\documentclass[intlimits,twoside,a4paper]{article}

\usepackage{amsmath,amssymb}
\usepackage{graphicx}
\usepackage{wrapfig}
\usepackage[T2A]{fontenc}
\usepackage[cp1251]{inputenc}

\usepackage{cmpj}

\issue{2011}{14}{2}{23601}

\doinumber{10.5488/CMP.14.23601}

\title[Effect of pressure on the electronic structure of hcp Titanium]%
{Effect of pressure on the electronic structure of hcp Titanium}
\author[M. Jafari \textsl{et al.}]
{M. Jafari\thanks{E-mail: jafari@kntu.ac.ir}\,, A. Jahandoost, Me. Vaezzadeh, N. Zarifi}
\address{Physics Department, K.N. Toosi University of Technology, Tehran, Iran}
\date{Received October 27, 2010, in final form March 11, 2011}

\authorcopyright{M. Jafari, A. Jahandoost, Me. Vaezzadeh, N. Zarifi, 2011}

\begin{document}

\maketitle

\begin{abstract}
The effect of pressure on the hexagonal close-packed structure of
titanium is investigated. The lattice parameters of the
equilibrium structure were determined in terms of the Gibbs free
energy using the Epitaxial Bain Path method. When this process was
repeated for several pressures, the effect of pressure on the
lattice parameters was revealed. The calculated lattice parameters
were in good agreement with the experimental and theoretical
results. The effects of pressure on parameters depending on the
electronic structure such as conductivity and resistivity in the
ground state were also investigated up to 30 GPa using density
functional theory.
\keywords effect of pressure, hcp-Ti, Gibbs free energy, lattice parameters, EBP method
\pacs 61.50.Ks, 05.70.Ce, 72.15.Eb, 71.15.Mb
\end{abstract}

\section{Introduction}

At room temperature and ambient pressure, Ti has a hexagonal
close-packed structure called the $\alpha$-phase. The lattice
parameters of this structure are $a=2.957$~{\AA}  and
$c=4.685$~{\AA}~\cite{1} in which the unit cell has two atoms at
$(1/3,\ 2/3,\ 1/4)$, $(2/3,\ 1/3,\ 3/4)$  and the space group
number is 194 ($P6_{3}/mmc$) with the $c/a$ ratio of
$\sim1.59$~\cite{2}. Experimental results at room temperature
indicate that lattice parameters decrease and the ${c}/{a}$ ratio
increases with pressure~\cite{3}. At ambient temperature and high
pressure, it changes to the $\omega $-phase~\cite{4,5}. The
lattice parameters of this structures are $a=4.598$~{\AA} and
$c=2.822$ ~{\AA}~\cite{1,6} with three atoms per unit cell at
$(0,\ 0,\ 0)$, $(1/3,\ 2/3,\ 1/2)$, $(2/3,\ 1/3,\ 1/2)$ and the
space group is ${P}6/{mmm}$ with the $c/a$ ratio of
$\sim0.61$~\cite{2}. The $\alpha\rightarrow\omega $ transition in
Titanium is a representative example of martensitic
transformation. Recently, Trinkel~et~al. have proposed two
pathways for this transformation which are called TAO-1
(``Titanium alpha to omega'') and TAO-2. This mechanism is a
direct mechanism in which six-atom transformation proceeds without
a meta-stable intermediate phase and has small shuffles and
strains~\cite{7}.

\looseness=-1The effect of pressure on lattice parameters was investigated
using the Epitaxial Bain Path (EBP) method~\cite{8,9}. In this
method, the equilibrium structure is determined by minimized Gibbs
free energy. As a hexagonal structure is defined by its lattice
parameters, the Gibbs free energy
[${G}\equiv{E}({a},{c})+{PV}({a},{c})$] should be minimized with
respect to ${a}$ and ${c}$. The EBP method is summarized and explained
here. At ${T}=0$~K, the Gibbs free energy was ${G}={E}+PV$, where
${P}$ is the pressure, ${V}$ was the volume per atom and ${E}$ is
the internal energy per atom. At ${P}\neq0$, the lattice parameter
of $a=a_{1}$ was chosen and the value of $c$ was varied until one
found the values $c=c_{1}$ and $E=E_{1}$\,. The energy of $E$ for
the lattice parameters ${c} = {c}_{i}$ (${i}=1,2,\dots$) with
constant ${a}_{1}$ was calculated using Wien2k until ${c}_{1}$ and
${E}_{1}$ values satisfied $\left({\partial E}/{\partial c}
\right)_{a} =-\left({Pa_{1}^{2} \sin \gamma }\right)/{2} $, where
$\gamma$ is the angle between ${a}$ and ${c}$. When a suitable
value of ${c}_{1}$ was determined, ${E}_{1}$\,, $V_{1}$ and the
Gibbs free energy ${G}_{1}$\,, which is ${G}_{1}={E}_{1}+{PV}_{1}$
 were found
at pressure ${P}$. This process was repeated for
several $a_{i}$ values at the same pressure and $G_{i}$ was
determined. For the equilibrium structure at pressure $P$, the
value giving a minimum $G_{\rm P}$ was chosen. Thus, by choosing
different values of $P$, the lattice parameters ${a}$ and ${c}$,
ratio ${c}/{a}$ and the volume were determined directly as the
functions of pressure.

The energy was calculated using Wien2k Package~\cite{10}, which
employs a self-consistent  Full-Potential Linearized Augmented
Plane wave plus local orbital (FLAPW+LO) method, under the
generalized gradient approximation (GGA) with the
Perdew-Burke-Emzerh of 96 exchange Correlation
functional~\cite{11}. Moreover, the following parameters were
used: Muffin-tin-radius, $RMT=2.3$~bohr, Largest vector in the
charge-density Fourier expansion, ${G}_{\rm max}=12$~bohr$^{-1}$,
${K}$ point $=4954$, Plane-wave cutoff, ${RK}_{\rm max}=9$, cutoff
energy $=-6$~Ryd and charge convergence $=1\times 10^{-4}\re$ (the
charge convergence was used in order to optimize parameters in the
SCF cycle). Except for $K$ point $=5599$ and $RMT=2.247$~bohr,
other parameters which were related to the effect of pressure on
the electronic structure did not change.

\section{Results and discussion}

\begin{figure}[h]
\includegraphics[width=0.47\textwidth]{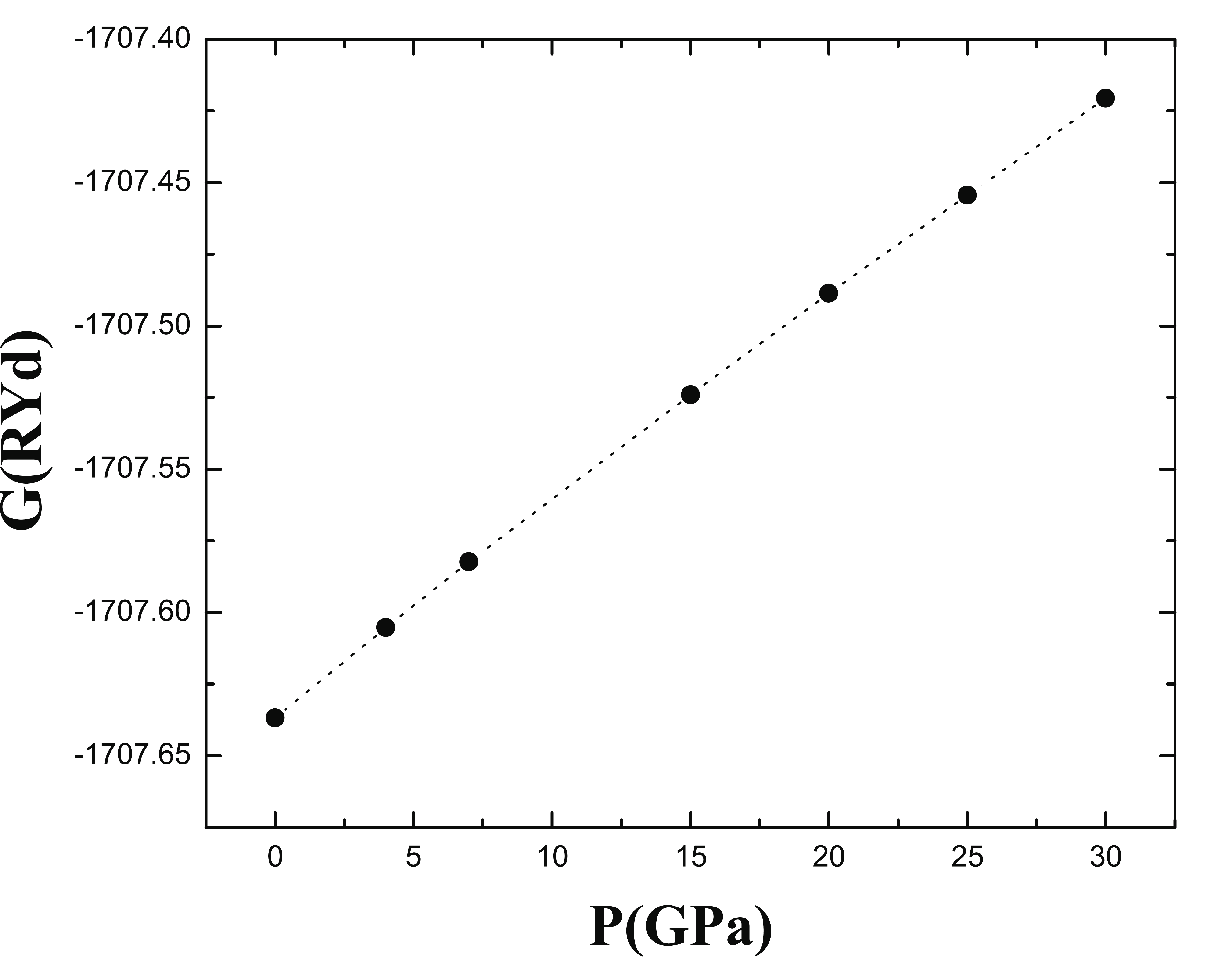}%
\hfill%
\includegraphics[width=0.45\textwidth]{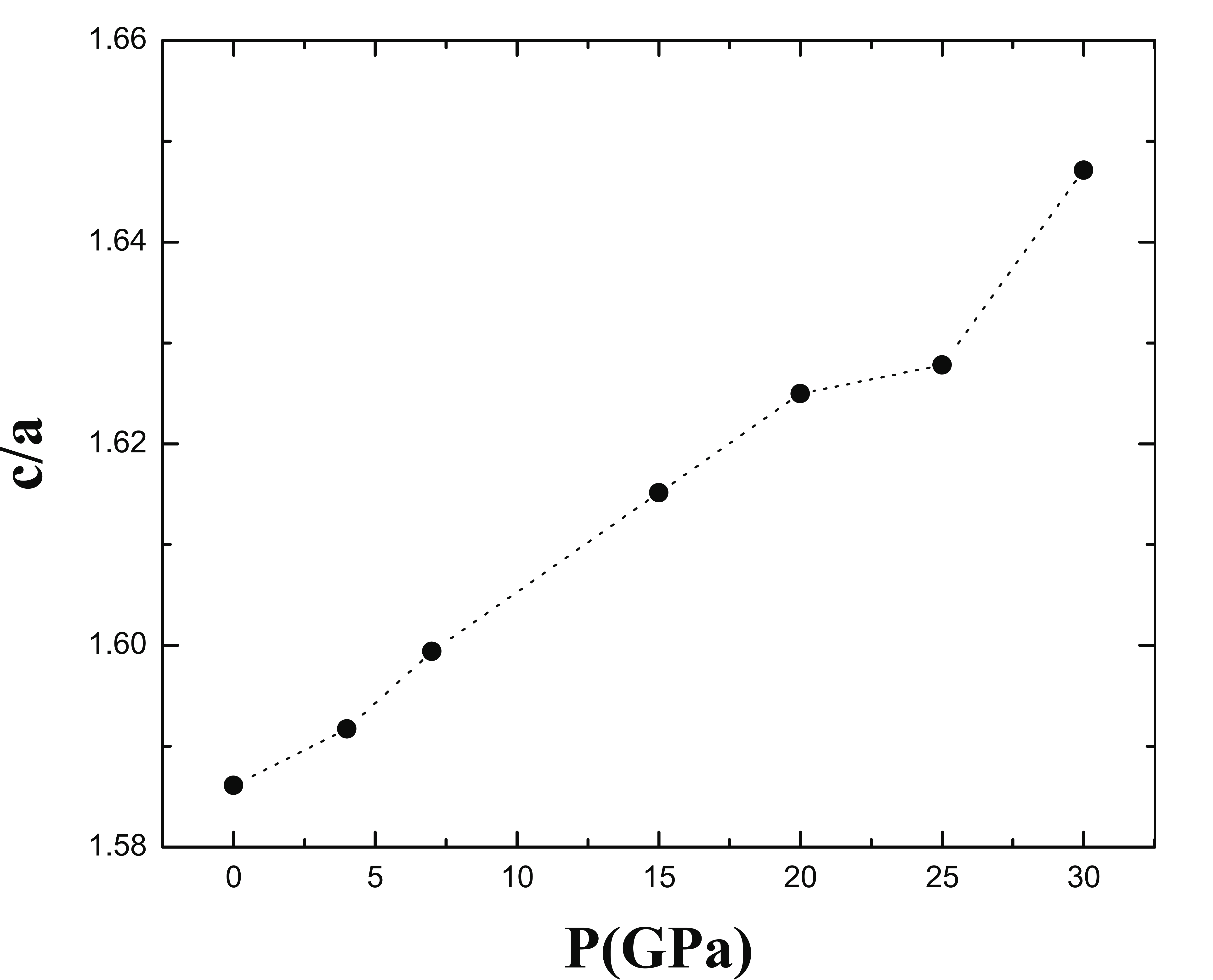}%
\\
\parbox[t]{0.48\textwidth}{%
\caption{The pressure dependence of the Free energy $G$.}
\label{fig1}}%
\hfill%
\parbox[t]{0.48\textwidth}{%
\caption{The pressure dependence of ratio $c/a$.} \label{fig2}}%
\end{figure}

\begin{figure}[h]
\hspace{0.35cm}\includegraphics[width=0.45\textwidth]{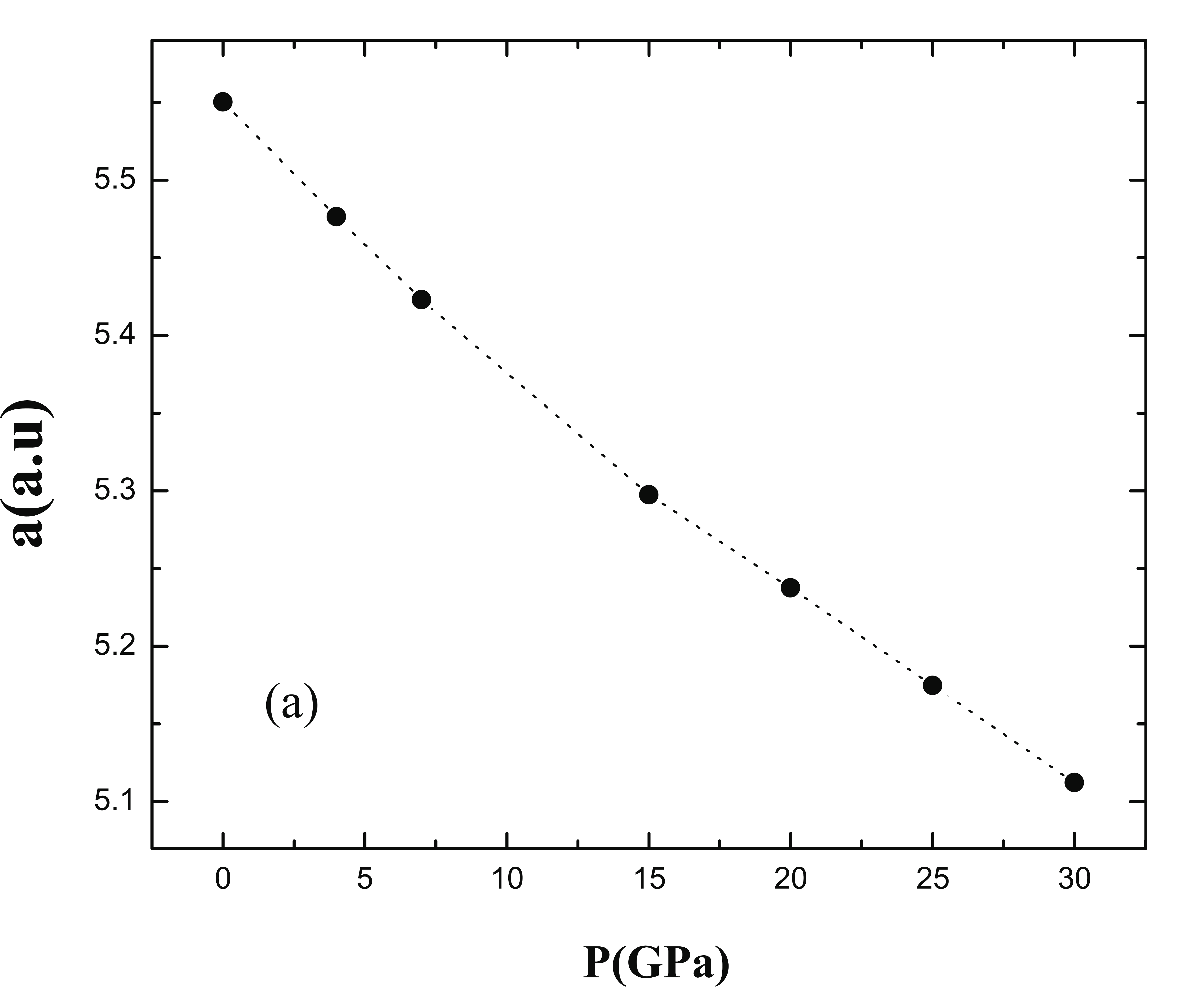}\hfill
\includegraphics[width=0.45\textwidth]{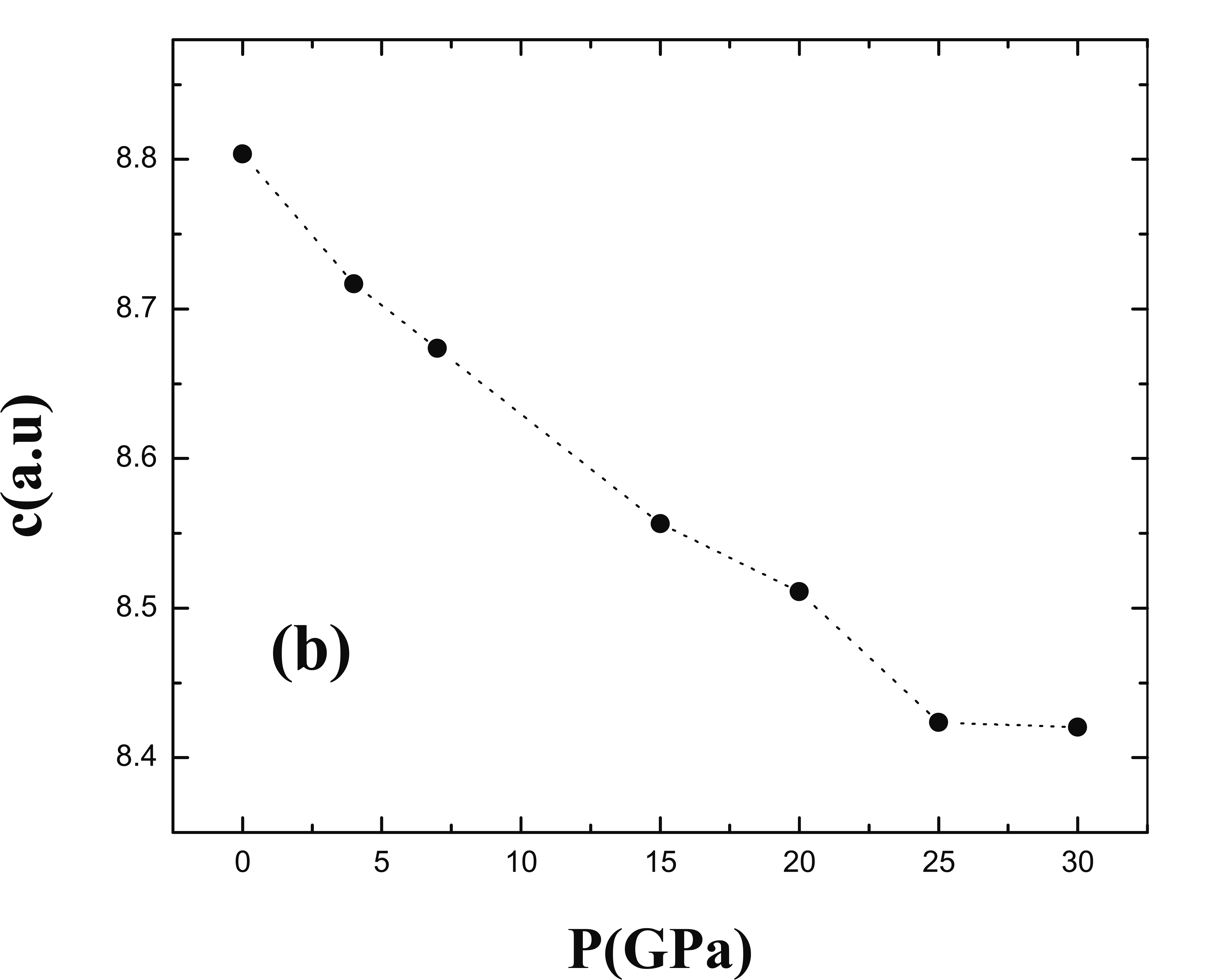}
\caption{The pressure dependence of (a) lattice parameter  $a$ and
(b) lattice parameter $c$.} \label{fig3}
\end{figure}

\begin{wrapfigure}{i}{0.5\textwidth}
\centerline {\includegraphics[width=0.45\textwidth]{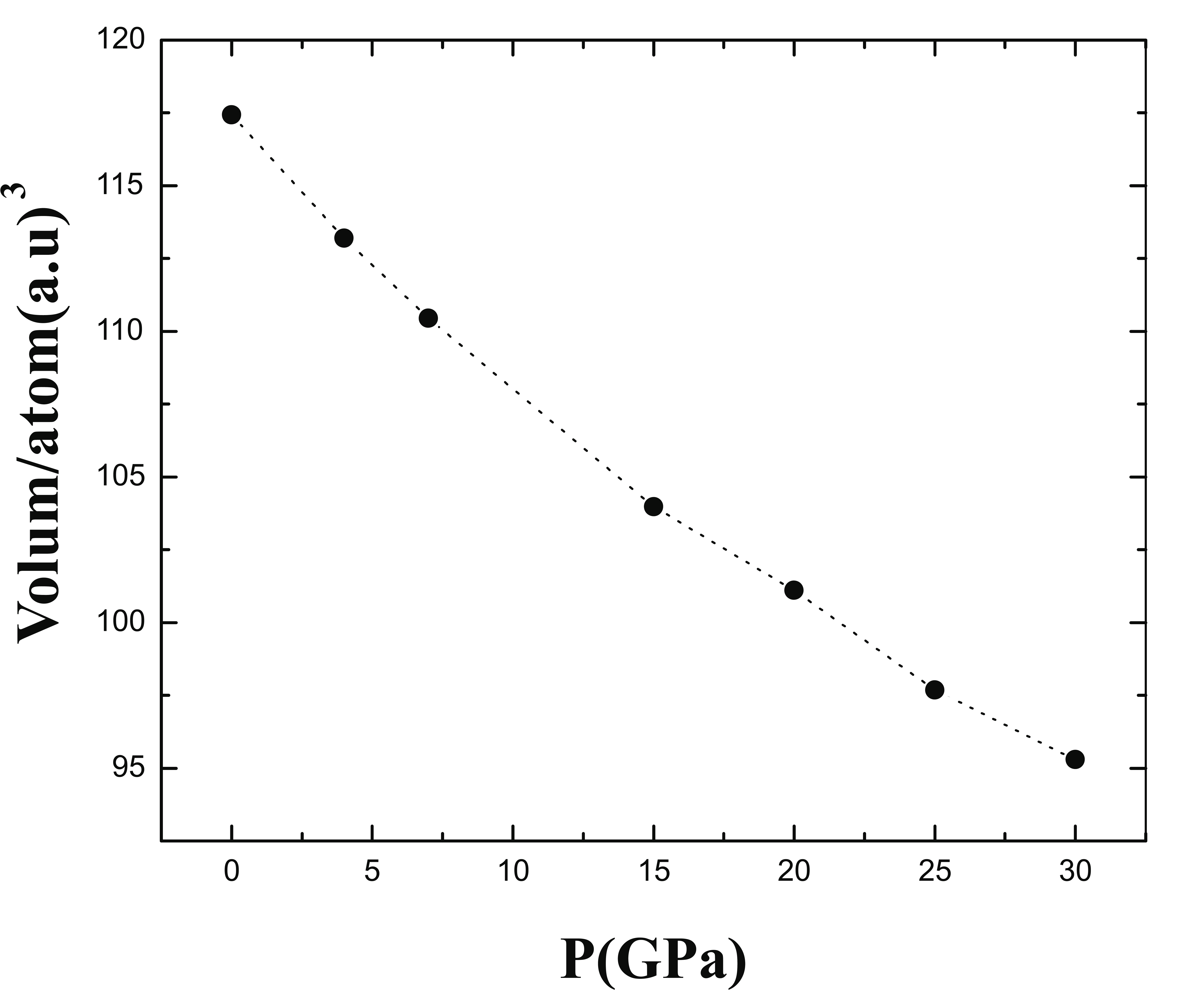}}
\caption{The pressure dependence of atomic volume.} \label{fig4}
\end{wrapfigure}

The effect of pressure on the hexagonal close-packed structure of
titanium in the ground state was investigated. The Gibbs free
energy (figure~\ref{fig1}), ratio of $c/{a}$ (figure~\ref{fig2}),
lattice parameters ${a}$ and ${c}$ (figure~\ref{fig3}) and volume
(figure~\ref{fig4}) were calculated as functions of pressure. The
experimental data were obtained using figure~5 in
Errandonea~et~al.~\cite{3} via Gate data software. These results
were in good agreement with theoretical~\cite{8} and experimental
studies~\cite{3}.


\begin{figure}[!h]
\includegraphics[width=0.45\textwidth]{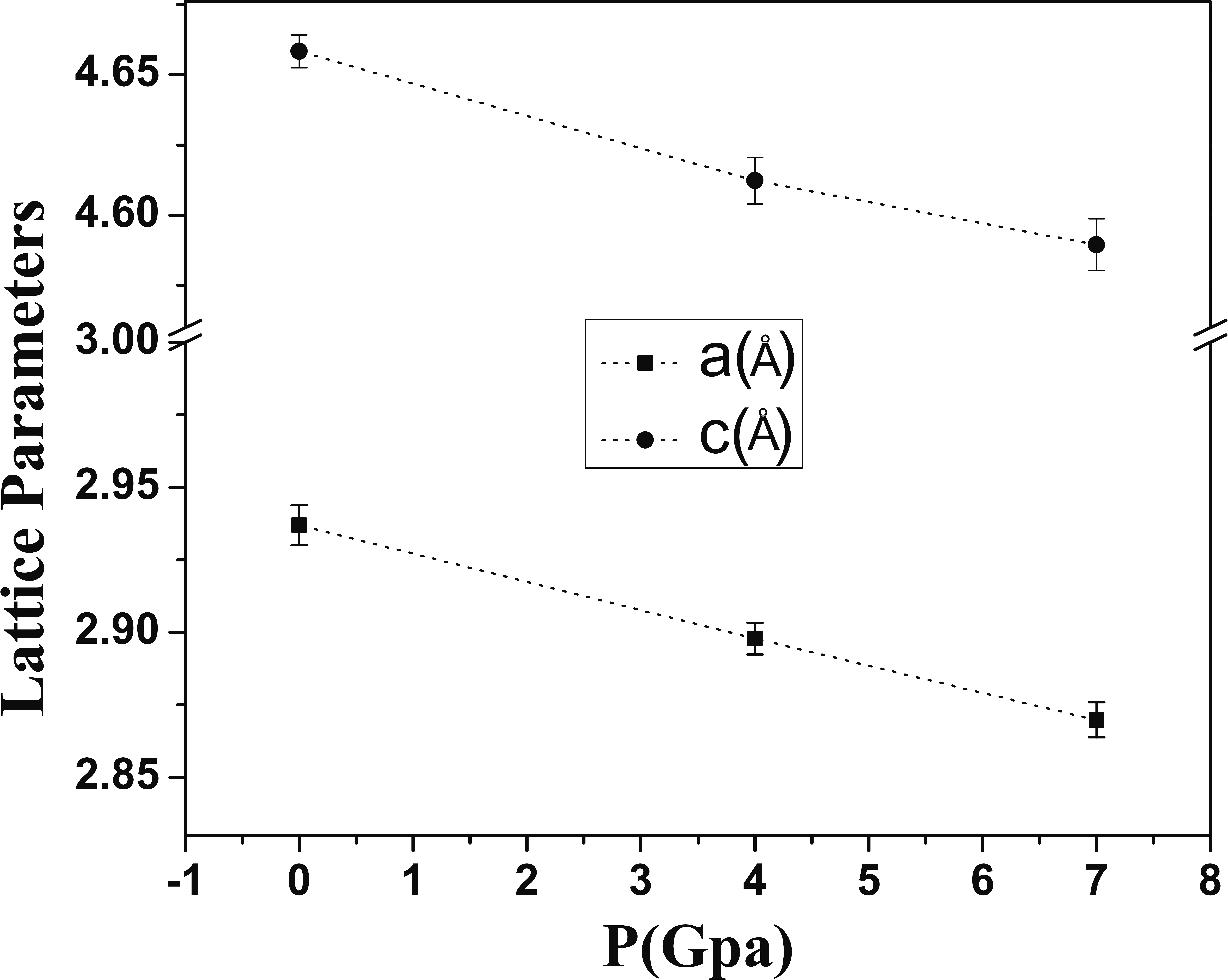}%
\hfill%
\includegraphics[width=0.435\textwidth]{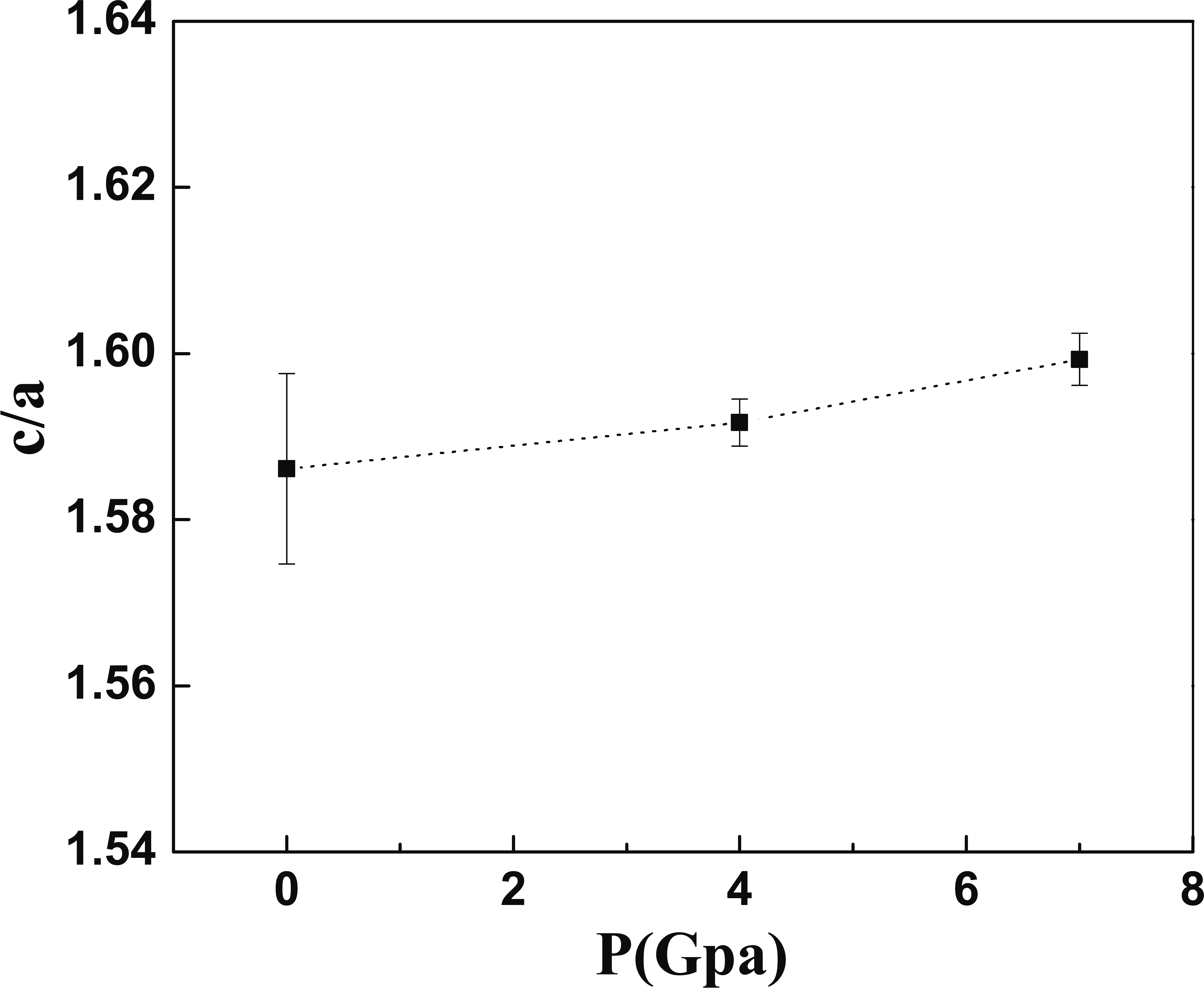}%
\\
\parbox[t]{0.48\textwidth}{%
\caption{Lattice  parameters vs  pressure for hcp Ti. Estimated
deviations with experimental work~\cite{3} are indicated.}
\label{fig5}}%
\hfill%
\parbox[t]{0.48\textwidth}{%
\caption{Ratio of $c/a$ parameters  vs   pressure for hcp Ti.
Estimated deviations with experimental work~\cite{3} are
indicated.} \label{fig6}}%
\end{figure}

In order to make a comparison between the results of this study
and those of the references~\cite{3}, only three points of the
pressure (0, 4 and 7~Gpa) were chosen which was due to the
limitation of the experimental results in which the experimental
points were only considered up to 14.5~GPa. These comparisons are
tabulated in tables~\ref{Tab1} to~\ref{Tab3} and shown in
figures~\ref{fig5} and~\ref{fig6}. As shown in the aforementioned
figures, based on the expectations, the lattice parameters
decreased and $c/a$ ratio increased with pressure.

In the present study at ${T}=0$~K and ${P}=0$~GPa, $\alpha $-phase
had the $c/a$ ratio of 1.586 which was in good agreement with the
experimental 1.584~\cite{3} and theoretical results
1.611~\cite{8}, 1.584~\cite{12} or 1.583~\cite{13} smaller than
the ideal value of 1.633 for an hcp crystal. Furthermore, the
experimental results showed that the $c/a$ ratio for $\alpha $-Ti
gradually increased  from 1.584 at atmospheric pressure to 1.622 at
14.5~GPa~\cite{3}.

\begin{table}[h]
\caption{Lattice parameter of hcp Ti at zero pressure; estimated
deviations with experimental work~\cite{3} are indicated.}
\label{Tab1} \vspace{2ex}
\begin{center}
\renewcommand{\arraystretch}{0}
\begin{tabular}{cccc} \hline
Lattice  parameter & This work  & Experimental$^a$~\cite{3} &
 Theoretical$^b$~\cite{8} \strut\\ \hline\hline
a (\AA) & 2.93693$\pm$0.007 & 2.9575 & 2.92455 \strut\\ \hline c
(\AA) & 4.65834$\pm$0.006 & 4.68548 & 4.71120 \strut\\ \hline c/a
& 1.58613$\pm$ 0.011 & 1.58427 & 1.61091 \strut\\ \hline
\end{tabular}
\renewcommand{\arraystretch}{1}
\end{center}
\end{table}

\begin{table}[h]
\caption{Lattice parameter of hcp Ti at 4~GPa; estimated
deviations with experimental work~\cite{3} are indicated.}
\label{Tab2} \vspace{2ex}
\begin{center}
\renewcommand{\arraystretch}{0}
\begin{tabular}{cccc} \hline
Lattice parameter & This work & Experimental$^{a}$~\cite{3} &
Theoretical$^{b}$~\cite{8} \strut \\ \hline\hline a (\AA) &
2.89777$\pm$0.005 & 2.9137 & 2.88898 \strut\\ \hline c (\AA) &
4.61230$\pm$0.008 & 4.65088 &4.65735 \strut\\ \hline
c/a   & 1.59167$\pm$0.003 & 1.59621 & 1.61211 \strut\\
\hline
\end{tabular}
\renewcommand{\arraystretch}{1}
\end{center}
\end{table}

\begin{table}[h]
\caption{Lattice parameter of hcp Ti at 7~GPa; estimated
deviations with experimental work~\cite{3} are indicated.}
\label{Tab3} \vspace{2ex}
\begin{center}
\renewcommand{\arraystretch}{0}
\begin{tabular}{cccc} \hline
Lattice parameter &  This work & Experimental$^{a}$~\cite{3} &
Theoretical$^{b}$~\cite{8} \strut \\ \hline\hline a (\AA) &
2.86973$\pm$0.006 & 2.88717 & 2.86523\strut \\ \hline
c (\AA) & 4.58955$\pm$0.009 & 4.63199 & 4.62227 \strut\\
\hline c/a   & 1.59930$\pm$0.003 & 1.60433 & 1.61323 \strut\\ \hline
\end{tabular}
\\\flushleft{\small \hspace{1cm}$^{a}$ Experimental values are obtained by Gate Data Software
from figure~5 of the reference~\cite{3}.\\
\hspace{1.cm}$^{b}$ Theoretical  values are obtained by Gate Data Software
from figure~2 of the reference~\cite{8}.}
\end{center}
\end{table}

The $\alpha $-phase can transform to the $\omega $-phase under
pressure. However, it is believed that both structures can coexist
in the pressure range studied here~\cite{5}. The effect of
pressure on lattice parameters of $\alpha $-phase in the range of
0--14.5~GPa was also investigated, both experimentally in~\cite{3}
and theoretically in~\cite{8}. These results confirm the
possibility of coexistence of both structures in the pressure
range of 2--9 GPa and even more.

In fact, the coexistence of these phases has been also reported by
experiments within the temperature range from room temperature to around 923~K~\cite{14} and
by the theory of reconstructive phase transitions~\cite{15}.

\begin{table}[!h]
\caption{Partial charges in the $s$, $p$ and $d$ bands.}
\label{Tab4} 
\begin{center}
\renewcommand{\arraystretch}{0}
\begin{tabular}{cccc} \hline
P (GPa) & $s$      & $p$      & $d$    \strut \\ \hline\hline
0 &  0.31515 & 0.2767  & 0.9858 \strut \\
4 & 0.32471 & 0.2870 & 2.0104 \strut \\
7 & 0.33159 & 0.2953 & 2.0426 \strut\\
15 & 0.34840 & 0.3125 & 2.0999 \strut\\
20 & 0.35676 & 0.3199 & 2.1242 \strut\\
25 & 0.36704 & 0.3315 & 2.1744 \strut\\
30 & 0.37429 & 0.3371 & 2.2127 \strut\\\hline
\end{tabular}
\renewcommand{\arraystretch}{1}
\end{center}
\vspace{-5mm}
\end{table}

\begin{table}[!h]
\parbox[t]{0.48\textwidth}{%
\caption{Density of states at the Fermi level as a function of
pressure.} \label{Tab5}}%
\hfill%
\parbox[t]{0.48\textwidth}{%
\caption{The pressure dependence of the Fermi energy.} \label{Tab6}}%
\\
\begin{center}
\renewcommand{\arraystretch}{0}
\begin{center}
\parbox[t]{0.48\textwidth}{%
\hspace{0.3cm}
\begin{tabular}{cc} \hline
P (GPa) & $n(E_{\rm F})$ (States/ev atom) \strut\\
\hline\hline 0 & 0.8930 \strut\\
  4 & 0.8868 \strut\\
  7 & 0.8500 \strut\\
  15 & 0.8625 \strut\\
  20 & 0.8000 \strut\\
  25 & 0.8158 \strut\\
  30 & 0.7908 \strut\\
\hline
\end{tabular}
}
\hfill%
\parbox[t]{0.48\textwidth}{%
\hspace{0.5cm}
\begin{tabular}{cc} \hline
P (GPa) & $E_{\rm F}$ (Ryd)\hspace{1cm} \strut\\
\hline\hline
  0 & 0.56228 \strut\\
  4 & 0.58999 \strut\\
  7 & 0.60881 \strut\\
  15 & 0.65693 \strut\\
  20 & 0.67996 \strut\\
  25 & 0.70925 \strut\\
  30 & 0.73021\strut\\
\hline
\end{tabular}
}
\end{center}
\renewcommand{\arraystretch}{1}
\end{center}
\end{table}

Using the lattice parameters and Wien2k, the number of electrons
in $s$, $p$ and $d$ bands (table~\ref{Tab4}), the Fermi energy
(table~\ref{Tab5}) and the density of states at this energy
${n}(\varepsilon_{\rm F})$ (table~\ref{Tab6}) were calculated for
different pressures.

\begin{wrapfigure}{i}{0.5\textwidth}
\centerline {\includegraphics[width=0.45\textwidth]{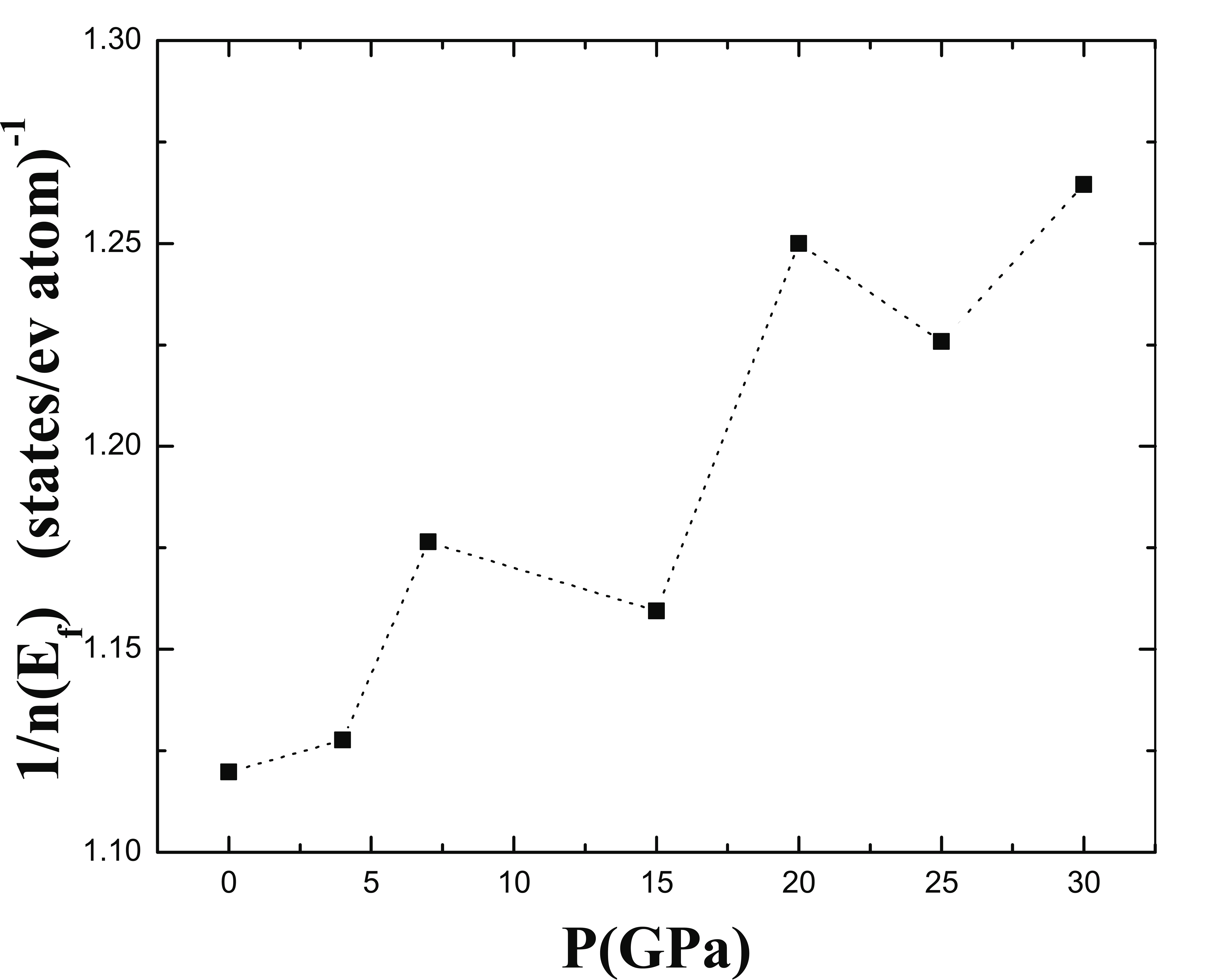}}
\caption{${1}/{n\left({\varepsilon}_f\right)}$ as a function of
pressure for hcp Ti.} \label{fig7}
\end{wrapfigure}

Moreover, the electrical conductivity can be expressed as  $\sigma
= \re^{2} \tau _{\rm F} v_{\rm F}^{2} n(\varepsilon _{\rm
F})/3$~\cite{16,17}, where ${v}_{\rm F}$ is velocity at the Fermi
energy and $\tau_{\rm F}$ is the relaxation time, but
${n}(\varepsilon_{\rm F})$ has a greater effect than the latter
two parameters. Figure~\ref{fig7} shows $1/{n}(\varepsilon_{\rm
F})$ as a function of pressure. Because $\rho\propto
1/{n}(\varepsilon_{\rm F})$, figure~\ref{fig7} can be taken as a
measure of the effect of pressure on electrical resistivity.

According to Matthiessen's rule, total electrical resistivity
$\rho$, due to electron scattering by different factors is given
by sum of these factors $\rho=\rho_{\rm Th}+\rho_{\rm D}+\rho_{\rm
I}$\,, where $\rho_{\rm Th}$ is thermal resistivity, $\rho_{\rm
D}$ and $\rho_{\rm I}$ are resistivity due to defects and
impurities, respectively. The other factor that can scatter
electrons is electron-electron (e-e) interaction, which is
negligible. At high temperature, the effects of impurities and
defects are negligible; thus, $\rho\approx\rho_{\rm Th}$\,; at low
temperature, $\rho_{\rm Th}$ is less than $\rho_{\rm D}+\rho_{\rm
I}$\,, so $\rho\approx\rho_{\rm D}+\rho_{\rm I}$\,. At ${T}=0$~K,
ions were frozen in fixed positions and electrons do not scatter
with phonon ($\rho_{\rm Th}=0$). In the present study, pure
titanium was investigated. It is chemically and thermodynamically
impossible to avoid impurities and defects, so the resistivity at
$T=0$ is not zero ($\rho=\rho_{\rm D}+\rho_{\rm I}\neq0$   where
$\rho =\rho_{\rm r}$ is called residual resistance).

Figure~\ref{fig7} shows that at ${T}=0$~K an increase in pressure
causes an increase in electrical resistivity, which contradicts
the results of P.S.~Balog, who investigated the phenomenon at
50--700$^\circ$C, where $\rho$ was due to electron-phonon
interaction. A pressure increase (at constant temperature) leads
to a decrease in inter-atomic spacing and atomic vibrational
amplitude, causing a decrease in electrical resistivity~\cite{18}.
However, in the present study, the investigation was carried out
at ${T}=0$~K and electrical resistivity increased with pressure.
According to table~\ref{Tab4}, an increase in pressure leads to an
increase in the number of electrons per volume.

Furthermore, according to table~\ref{Tab5}, an increase in
pressure leads to a decrease in the density of state, thus causing an
increase in the electrical resistivity.

\looseness=-1According to table~\ref{Tab6}, an increase in pressure leads to an
increase in the width of the valence band. In the formation of
molecules, several atoms are arranged beside each other, so atomic
orbitals are split and several molecular orbitals are created
while the number of orbitals is proportional to the number of
atoms. However, if the inter-atomic space is small, atomic orbital
splitting is larger. According to the calculations and
figure~\ref{fig2}, a pressure increase causes a decrease in the
lattice constant and inter-atomic space, thus increasing the
orbital splitting and width of the valence band.
\begin{table}[h]
 \vspace{-2ex}
\caption{Partial charges in the $p_x+p_y\,,p_z$\,;
$d_{z^2}\,,d_{xy}+d_{x^2-y^2}$ and $d_{xz}+d_{yz}$\,.}
\label{Tab7}
\begin{center}
\renewcommand{\arraystretch}{0}
\begin{tabular}{cccccc} \hline
P (GPa) & $p_x+p_y$ & $p_z$ & $d_{xy}+d_{x^2-y^2}$ & $d_{z^2}$ &
$d_{xz}+d_{yz}$ \strut\\\hline\hline %
 0 & 0.18896 & 0.08777 & 0.84783 & 0.46065 & 0.67736 \strut\\
 4 & 0.19654 & 0.09052 & 0.86045 & 0.46494 & 0.68507 \strut\\
 7 & 0.20200 & 0.09334 & 0.87419 & 0.47196 & 0.69646 \strut\\ %
15 & 0.21359 & 0.09895 & 0.90120 & 0.48449 & 0.71430 \strut\\ %
20 & 0.21833 & 0.10159 & 0.91282 & 0.48769 & 0.72351 \strut\\ %
25 & 0.22668 & 0.10482 & 0.93479 & 0.50003 & 0.73965 \strut\\ %
30 & 0.22877 & 0.10840 & 0.95242 & 0.50818 & 0.75218 \strut\\\hline %
\end{tabular}
\renewcommand{\arraystretch}{1}
\end{center}
\end{table}

Table~\ref{Tab7} lists the number of electrons in $s$, $p$ and $d$
orbitals and the deviation from spherical symmetry is shown in
table~\ref{Tab8}. This deviation for $p$ and $d$ orbitals is given
by~\cite{19}:
\[
\Delta n_{{d}} =\left( n_{{d}_{xy}} +n_{{d}_{{x^2}-{y^2}}
}\right)-\frac{1}{2} \left(n_{{d}_{xz}}+n_{{d}_{{yz}} }
\right)-n_{{d}_{{z^2}}}\,,
\]
\[
\Delta n_{{p}} =\frac{1}{2} \left(n_{{p}_{{x}} } +n_{{p}_{{y}} }
\right)-n_{{p}_{{z}}}\,.
\]

\begin{table}[h]
\caption{Deviation from spherical symmetry of the $p$ and $d$
states as a function of pressure.} \label{Tab8} 
\begin{center}
\renewcommand{\arraystretch}{0}
\begin{tabular}{ccc} \hline
P (GPa) & $\Delta n_{d}$ & $\Delta n_{p}$ \strut\\ \hline\hline %
 0 & 0.048 & 2.1528 \strut\\ %
 4 & 0.052 & 2.1712 \strut\\ %
 7 & 0.054 & 2.1641 \strut\\ %
15 & 0.059 & 2.1585 \strut\\ %
20 & 0.063 & 2.1491 \strut\\ %
25 & 0.064 & 2.1625 \strut\\ %
30 & 0.068 & 2.1104 \strut\\ \hline%
\end{tabular}
\renewcommand{\arraystretch}{1}
\end{center}
\end{table}
If $\Delta{n}_{p}$ and $\Delta{n}_{d}$ are close to zero,
deviation from spherical symmetry will be just slight. According to
table~\ref{Tab8}, this deviation increases with pressure for $d$
orbitals.

\section{Conclusion}

The aim of the present study was to investigate the pressure effect on
lattice parameters of hcp structure in titanium. The obtained results
showed that the $c/a$ ratio of hcp was nearly constant. However,
it is believed that both structures can coexist in the
pressure range studied. The alpha phase was the most stable phase at
ambient conditions and its transformation to the omega phase in
the pressure range of 2--9~GPa. Moreover, theoretical and
experimental results confirmed the possibility of coexistence of
both structures within the pressure range of 2--9~GPa and even more.

Furthermore, effects of pressure on parameters depending on the
electronic structure, such as conductivity, resistivity, the Fermi
energy and $n(\varepsilon_{\rm F})$ in the ground state were also
investigated up to 30~GPa using density functional theory.
Moreover, an increase in pressure leads to a decrease in the density
of state, thus causing an increase in the electrical resistivity.

\ukrainianpart

\title{Вплив тиску на електронну структуру hcp титану}

\author{М. Джафарі, А. Джагандуст, М. Ваеззаде, Н. Заріфі}

\address{Технологічний університет ім. К.Н. Тусі, Тегеран, Іран}

\makeukrtitle

\begin{abstract}
Досліджено вплив тиску на гексагональну щільно упаковану структуру титану.
Параметри ґратки визначалися  в термінах вільної енергії Ґіббса, використовуючи
метод епітаксії шляхом Бейна. Коли цей процес повторювався для декількох тисків, було виявлено вплив тиску  на параметри ґратки.
Обчислені параметри ґратки добре узгоджувалися із теоретичними та експериментальними результатами. Вплив тиску на параметри, такі як провідність та опір в основному стані, в залежності від електронної структури, також було досліджено аж до 30~GPa, використовуючи
теорію функціоналу густини.
\keywords вплив тиску, hcp-Ti, вільна енергія Ґіббса, параметри ґратки, метод EBP

\end{abstract}
\end{document}